# Optical frequency comb Fourier transform spectroscopy of formaldehyde in the 1250 to 1390 cm$^{-1}$ range: experimental line list and improved MARVEL analysis

Matthias Germann[1], Adrian Hjältén[1], Jonathan Tennyson[2], Sergei N. Yurchenko[2], Iouli E. Gordon[3], Christian Pett[4], Isak Silander[1], Karol Krzempek[5], Arkadiusz Hudzikowski[5], Aleksander Głuszek[5], Grzegorz Soboń[5], and Aleksandra Foltynowicz[1,+]

[1] *Department of Physics, Umeå University, 901 87 Umeå, Sweden*
[2] *Department of Physics and Astronomy, University College London, Gower Street, London WC1E 6BT, United Kingdom*
[3] *Center for Astrophysics, Harvard & Smithsonian, Atomic and Molecular Physics Division, Cambridge, MA 02138, US*
[4] *Department of Chemistry, Umeå University, 901 87 Umeå, Sweden*
[5] *Laser & Fiber Electronics Group, Faculty of Electronics, Photonics and Microsystems, Wrocław University of Science and Technology, Wybrzeże Wyspiańskiego 27, 50-370 Wrocław, Poland*
[+]*aleksandra.foltynowicz@umu.se*

We use optical frequency comb Fourier transform spectroscopy to record high-resolution, low-pressure, room-temperature spectra of formaldehyde (H$_2^{12}$C$^{16}$O) in the range of 1250 to 1390 cm$^{-1}$. Through line-by-line fitting, we retrieve line positions and intensities of 747 rovibrational transitions: 558 from the $\nu_6$ band, 129 from the $\nu_4$ band, and 14 from the $\nu_3$ band, as well as 46 from four different hot bands. We incorporate the accurate and precise line positions (0.4 MHz median uncertainty) into the MARVEL (measured active vibration-rotation energy levels) analysis of the H$_2$CO spectrum. This increases the number of MARVEL-predicted energy levels by 82 and of rovibrational transitions by 5382, and substantially reduces uncertainties of MARVEL-derived H$_2$CO energy levels over a large range: from pure rotational levels below 200 cm$^{-1}$ up to multiply excited vibrational levels at 6000 cm$^{-1}$. This work is an important step toward filling the gaps in formaldehyde data in the HITRAN database.

## 1 Introduction

Formaldehyde, H$_2$CO, is an important chemical species for both applied and fundamental science. The relevance of formaldehyde extends from astrophysics, industrial applications and environmental monitoring, through public health and toxicology to fundamental chemistry and physics.

For astrophysics, formaldehyde plays a crucial role as an abundant constituent of the interstellar medium[1,2]. It is regarded as a precursor for the formation of complex organic molecules such as amino acids[3]. Galactic and extragalactic interstellar formaldehyde masers have been found, which have been utilized as tracers for high-density environments such as star-forming regions in galaxies (see [4] and references therein). Formaldehyde has been found in several comets, e.g., Halley and Hale-Bopp (see [5] and references therein) and in proto-planetary discs around young low-mass stars (see, e.g., [6] and references in [7]). Possible future detection of formaldehyde on an exoplanet might draw particular attention due to its role as a potential biomarker according to the RNA world hypothesis (see [7] and references therein). Finally, formaldehyde absorption lines of cosmologically distant absorbers in front of a quasar have been used, in conjunction with ammonia and carbon monosulfide lines, to constrain a putative variation of the proton-electron mass ratio on cosmological time scales[8].

Formaldehyde is not only abundant in outer space but also on Earth and in our everyday life. With an annual world production of over twenty million tons, formaldehyde is a crucial substance for the chemical industry. It is used (among other applications) to produce formaldehyde resins or as an intermediate in the production of other industrial chemicals[9, 10]. Formaldehyde-based resins are frequently used in construction materials. Together with its adverse health effects – $H_2CO$ is allergenic, toxic, and carcinogenic – this makes formaldehyde an important indoor air pollutant[9-11].

Outdoors, formaldehyde is also one of the most abundant hydrocarbons in the atmosphere[12]. It is emitted from biomass burning and fossil fuel combustion, as well as produced in situ through oxidation of hydrocarbons, such as methane and other volatile organic compounds (VOCs)[12-14]. Atmospheric formaldehyde is readily decomposed through photolysis and oxidation by OH radicals and thus has a short lifetime, on the order of hours[12, 14]. Therefore, atmospheric formaldehyde has been used as a proxy for emissions of non-methane VOCs[12, 14] and for tropospheric OH[15].

In fundamental science, formaldehyde is the simplest aldehyde and as such a prominent example of this class of compounds in chemistry. For molecular physics and spectroscopy, formaldehyde represents a relatively simple, highly symmetric (point group $C_{2v}$) asymmetric top, exemplifying effects such as Coriolis couplings between vibrational bands. Historically, formaldehyde has thus amassed several "firsts": it was among the first molecules to be studied with microwave spectroscopy (see [5] and references therein) and the first polyatomic molecule for which the rotational structure of vibronic transitions was understood[16]. A particularly interesting feature of formaldehyde, at the interface between molecular physics and physical chemistry, is its electronic ground-state potential energy surface (PES). Besides the local minimum corresponding to the structure of formaldehyde, the PES has additional local minima corresponding to cis- and trans-hydroxycarbene[17, 18] – where for the latter tunneling to the formaldehyde minimum has been observed[19] – as well as asymptotic regions corresponding to the $CO \cdot H_2$ complex[20, 21].

Owing to this wide relevance of spectroscopic investigation, detection, and monitoring of formaldehyde, the $H_2CO$ rovibrational spectrum has been studied extensively. The spectroscopic data from various experimental techniques have been compiled and (re-)analyzed several times over the years, e.g., by Clouthier et al.[22], Perrin et al.[23] and Kwabia Tchana et al.[24]. Most recently, Al-Derzi et al.[25] performed a comprehensive analysis of the available infrared spectroscopic data for formaldehyde. They validated more than ten thousand non-redundant transitions from 43 sources of high-resolution measurements and used the MARVEL procedure[26-28] to provide 5029 empirically determined energy levels up to 6188 cm$^{-1}$ above the rovibrational ground state. They used these energy levels to update the ExoMol/AYTY line list for hot $H_2CO$ computed by Al-Refaie et al.[7] (based on an empirically refined *ab initio* PES and an *ab initio* dipole moment surface), which resulted in a line list with 367 778 transition frequencies, between $9.5 \times 10^{-7}$ cm$^{-1}$ and 5887 cm$^{-1}$, determined with experimental accuracy. Below, we refer to this line list as the MARVELized ExoMol/AYTY line list to distinguish it from the earlier calculated line list from Al-Refaie et al.[7], which we will refer to as the semi-empirical ExoMol/AYTY line list.

Despite these efforts, major spectroscopic databases, such as HITRAN, still have considerable gaps in the mid-infrared (MIR) spectrum of formaldehyde. In the HITRAN2020 edition[29], $H_2CO$ infrared absorption at wavelengths from about 6 to 20 µm (510 to 1620 cm$^{-1}$) is missing. Until now, satellite-based atmospheric monitoring of formaldehyde has been largely carried out in the microwave[30] and UV[31] regions, while in the MIR only the 3.6 µm[32] and 5.7 µm[33] bands have been used. Tunable-laser-based sensing at atmospheric conditions and in combustion environments has also been performed mainly on absorption lines in the 3.6 µm range[34-39] and 5.7 µm range[40, 41], reaching part-per-billion level[42] and part-per-trillion[40, 43] level sensitivities. The $\nu_4$, $\nu_6$ band region around 8 µm has not received as much attention in terms of sensing and line list compilation because absorption in this region is about an order of magnitude weaker than that in the 3.6 and 5.7 µm bands. Moreover, these bands overlap with methane and (to a lesser extent) water vapor bands that dominate in this spectral region. However, the sensitivity and resolution of the modern instruments have substantially improved in recent years and therefore accurate line lists of these bands are needed from the perspective of atmospheric remote sensing. For example, the IASI-NG instrument (Infrared Atmospheric Sounding Interferometer - New Generation) onboard of EUMETSAT's EPS-SG satellites covers the 8 µm range and achieves a higher spectral resolution and lower noise than its IASI predecessor[44]. Moreover, the 8 µm range is within the sensitivity of the Mid-Infrared Instrument (MIRI) onboard the recently launched James Webb Space Telescope[45], rendering these wavelengths well suited for astrophysical observations.

The formaldehyde spectrum in the 8 µm spectral range has been studied with conventional Fourier transform infrared spectroscopy (FTIR) by Allegrini et al.[46], Nakanaga et al.[47] and Nadler et al.[48], and with FTIR and tunable diode laser spectroscopy by Reuter et al.[49]. Microwave and millimeter-wave spectra of rotational transitions within vibrational excited states, as those from Oka et al.[50], Dangoisse et al.[51] and Margulès et al.[52], complement the FTIR measurements by providing highly accurate spacings between rotational levels from the vibrational excited states of bands in the 8 µm range. As mentioned in these studies[49, 52], these bands are extremely perturbed making their analyses and modelling for predictions complex. The perturbations are also harder to represent in the *ab initio* calculations which, in particular, can lead to large changes in predicted intensities. We note that the MARVEL procedure used in this work is model-independent and the accuracy of the resulting energy levels is not affected by perturbations. The previous experimental works report spectroscopic constants including those for deperturbation analyses, however without having the programs that were used in the fits, one cannot reproduce the spectrum with openly available community tools, including PGOPHER[53].

Here we make use of the high resolution, absolute frequency calibration, large bandwidth, and high spectral brightness of a recently developed optical frequency comb Fourier transform spectrometer[54, 55] to record formaldehyde absorption spectra in the range from 1250 to 1390 cm$^{-1}$ (7.2 to 8 µm). Through line-by-line fitting, we retrieve the positions and intensities of 747 absorption lines from three fundamental bands ($\nu_3$, $\nu_4$, $\nu_6$) and four hot bands ($2\nu_6 - \nu_6$, $2\nu_6 - \nu_4$, $\nu_4 + \nu_6 - \nu_4$ and $\nu_3 + \nu_6 - \nu_3$) of $H_2^{12}C^{16}O$, with a median frequency uncertainty of 0.4 MHz and a relative intensity uncertainty of 15%. We include these accurate line positions into the MARVEL analysis of the rotational and rovibrational formaldehyde spectrum, which results in 82 new MARVEL-derived energy levels and increases the number of MARVEL-predicted transitions by 5382. For levels already treated by the MARVEL analysis and directly probed here through spectroscopic transitions, the uncertainties are reduced by up to two orders of magnitude. Significant uncertainty reductions are also achieved for levels not directly probed here – from below 100 cm$^{-1}$ up to 6000 cm$^{-1}$ – via so-called spectroscopic networks exploited in the MARVEL analysis. Of particular interest here are multiply vibrationally excited levels giving rise to overtone and combination bands within the sensitivity range of several instruments at the James Webb Space Telescope (FGS-NIRISS, NIRcam, NIRSpec)[56] and of the upcoming ARIEL (Atmospheric Remote-sensing Infrared Exoplanet Large-survey) mission[57].

## 2 Experimental setup and procedures

### 2.1 *Spectrometer and measurement procedure*

The spectrometer setup has been described in detail previously in Refs. [54, 55]. Its main components are an optical frequency comb source, a multi-pass absorption gas cell and a Fourier transform spectrometer (FTS).

The comb source is a compact fiber-based system described in Ref. [58], which produces an 8 µm comb through difference frequency generation in an orientation-patterned gallium phosphide (OP-GaP) crystal with a poling period of 58 µm. The pump and signal combs are derived from a common femtosecond Er:fiber oscillator with a repetition frequency ($f_{rep}$) of 125 MHz. Because of this, the 8 µm idler comb is free from carrier envelope offset, $f_{ceo}$, and is thus fully stabilized via the control of $f_{rep}$ using a fiber stretcher in the oscillator cavity. The $f_{rep}$ is phase-locked to a tunable radio frequency generator referenced to a GPS-disciplined rubidium clock.

The 8 µm comb beam, with a power of ~1 mW, is collimated and coupled into a Herriott-type multi-pass cell (Thorlabs, HC10L/M-M02) with a path length of 10.436(15) m, filled with the formaldehyde sample gas. Two pressure transducers connected to the multi-pass cell are used to continuously monitor the cell pressure: a CERAVAC model CTR 100 N 1 (resolution 0.0001 mbar) and a CERAVAC CTR 101 N 100 (resolution 0.01 mbar) for pressures below and above 1 Torr (1.3 mbar), respectively. The cell temperature is measured using a Pt100 resistance temperature detector mounted on the outer glass surface of the cell.

Behind the multi-pass cell, the comb beam is recollimated and directed to a home-built fast-scanning FTS. The two out-of-phase FTS outputs are detected with two HgCdTe detectors in a balanced configuration. Two irises in front of the detectors are used to balance the incident power levels and to keep them within the linear response range of the detectors. The differential signal between the detectors is digitized and recorded using a custom LabVIEW program. The optical path difference (OPD) of the FTS is calibrated using a stable 1.56 µm continuous-wave (CW) diode laser propagating on a path parallel to the comb beam. To eliminate the influence of dispersion of ambient H$_2$O on the reference laser wavelength calibration, the FTS is purged with dry air during the measurements.

We recorded formaldehyde spectra using the sub-nominal sampling-interleaving technique described in Refs. [59, 60]. At the start of the absorption measurement, we stabilized $f_{rep}$ and acquired 50 interferograms. We then stepped $f_{rep}$ 10 times at 40 Hz increments, corresponding to 12.5 MHz steps of the comb lines in the optical domain, and acquired 50 interferograms at each step. We repeated this process ten times stepping $f_{rep}$ in alternating directions to produce a total of 400 interferograms at each $f_{rep}$ step. One interferogram was acquired in 3.7 s, resulting in a total acquisition time of 4.1 h. Prior to the absorption measurement, we acquired 400 reference interferograms with the cell evacuated and $f_{rep}$ locked to the value of the first step, for later normalization of the absorption spectra.

To optimize the signal-to-noise ratio for lines with intensities spanning more than three orders of magnitude, we acquired formaldehyde spectra in four separate measurement runs, each of them recorded at a different formaldehyde pressure and on a different day over the course of about three months. The experimental conditions for these four runs are summarized in Table 1, where they are referred to as run (1) to (4) in the order of increasing pressure, corresponding also to the chronological order of the measurements.

Table 1: Measurement conditions and number of analyzed lines for each measurement run. Columns: 'Measurement run', designation of the respective measurement run. 'Pressure', approximate formaldehyde pressure in the absorption cell[a]. 'Temperature', average temperature of the absorption cell during the measurement. 'Lines fitted', number of absorption lines whose positions and intensities have been retrieved by line fitting. 'Lines assigned', number of fitted lines that have been assigned to formaldehyde transitions, separated into those with and without MARVEL predicted line positions.

| Measurement run | Pressure [mbar] | Temperature [°C] | Lines fitted | Lines assigned | |
|---|---|---|---|---|---|
| | | | | MARVEL-predicted | Non-MARVEL |
| (1) | 0.02 | 21.7 | 336 | 336 | – |
| (2) | 0.12 | 21.9 | 354 | 308 | 30 |
| (3) | 0.58 | 21.8 | 358 | 162 | 58 |
| (4) | 2.47 | 21.8 | 464 | 151 | 67 |

## 2.2 Sample preparation

Formaldehyde gas samples were prepared through pyrolysis of solid paraformaldehyde, i.e., polymerized formaldehyde[61]. Paraformaldehyde granulate was heated under vacuum to temperatures above 120 °C (see below for preparation details of individual measurements). The formed formaldehyde vapor was guided through a cold trap placed in a dry-ice cooling bath to capture impurities. Monomeric formaldehyde was collected in a sample container placed in a second cooling bath at cryogenic temperatures.

We began the sample preparation by filling a few grams of paraformaldehyde granulate (Alfa Aesar, 97%, CAS number 30525-89-4, natural isotopic abundances) in a round bottom flask. We attached the flask to the pyrolysis setup, flushed the setup a few times with nitrogen and evacuated it to a pressure ≤ 2 mbar. We then sealed the setup from the pump by closing a valve in the pump line and heated the flask with a heating mantle. When the paraformaldehyde temperature reached about 90 °C, pyrolysis set in as noticed by a distinct increase in the pressure. We continued the pyrolysis for a few minutes until we saw a clear decrease in the amount of paraformaldehyde granulate left. We then sealed off the sample container with a valve and disconnected it from the rest of the pyrolysis setup.

---

[a] The values shown are the time averages of the pressure over the period of acquisition of the spectra for each run. The pressure values from run (1) and (4) have been corrected for the cell leak rate retrieved from run (1), see Sec. 3.2.2 for details. The pressure values from run (2) and (3) are indicative only, i.e., are not used in the line intensity calculations.

For safety reasons, the pyrolysis was performed in a fume hood located a few meters away from the spectrometer. Next, we thus took the sample container, immersed in the cold bath, to the spectrometer, where we attached it to its gas supply system. We flushed the gas line several times with nitrogen, then evacuated it and filled the multi-pass absorption cell (which was continuously pumped between the measurements): we carefully lowered the Dewar with the cold bath to expose a part of the sample container to ambient temperature while monitoring the resulting increase in the gas pressure. If the initial filling resulted in a cell pressure above the desired one, we pumped the cell until the desired pressure was reached.

Gaseous formaldehyde is known to be unstable – polymerization and adsorption on surfaces result in a decrease of the formaldehyde partial pressure (see, e.g., Ref. [47, 62]). After filling the cell, we therefore waited about one hour before recording spectra, for the formaldehyde pressure to settle.

Over the course of the experimental work for this study, we steadily improved the sample preparation process. For runs (1) and (2), we immersed the cold trap in a dry-ice/acetone bath and the sample collector in a liquid nitrogen bath. For run (3), we used a dry-ice/isopropanol bath for the cooling trap (as to avoid the highly volatile acetone while reaching practically the same temperature) and an ethanol "slosh bath", i.e., a mixture of solid and liquid ethanol, for the sample collector, thus increasing the temperature from the nitrogen boiling point, 77 K, to ~157 K [63]. Being above the boiling point of oxygen (90 K), this temperature prevents condensation of oxygen from ambient air possibly leaking into the setup. Furthermore, we reduced the maximum temperature used for pyrolysis from >200 °C for run (1) to <160 °C and <130 °C for runs (2) and (3), respectively. For run (1), we used a glass vial with a plastic cap as a sample collector. The vial was connected to the pyrolysis setup with a stainless-steel tube protruding into the vial through a tight hole in the plastic cap. For runs (2) and (3), we instead used a piece of stainless-steel tube, fused at one end by welding, as a sample collector, thus avoiding any potential leaks at the steel-plastic and plastic-glass interfaces. The gas sample for run (4) was taken from the formaldehyde sample prepared for run (3), which we prepared on the previous day and stored overnight immersed in an ethanol cold bath.

## 2.3 *Spectral treatment*

We processed the raw data from the spectrometer using custom MATLAB scripts in the way described in Ref. [55]. We took the Fourier transform of the interferograms and averaged the 400 spectra recorded at each $f_{\text{rep}}$ step, after matching the sampling points to the comb line frequencies as described in Refs. [59, 60]. This requires determining the effective wavelength, $\lambda_{\text{ref}}$, of the CW reference laser used for OPD calibration, which depends on the alignment and collimation of the comb beam relative to the CW laser beam. This effective reference wavelength is not known a priori with sufficient accuracy, but can be found empirically by minimizing instrumental line shape distortions in the spectra[60]. As observed previously[55], the optimum value of $\lambda_{\text{ref}}$ varied across the spectrum. To correct for this variation we divided each spectrum into three to six segments (depending on the observed variation) and determined an effective $\lambda_{\text{ref}}$ for each segment as described in Appendix A of Ref. [55]. We then normalized the frequency-calibrated spectra to the averaged reference spectrum and converted the resulting transmission spectra to absorption spectra using the Lambert-Beer law. To correct the baseline in the absorption spectra, we fitted a model consisting of a sum of a simulated absorption spectrum, based on the MARVELized ExoMol/AYTY line list, and a baseline consisting of a $5^{th}$ order polynomial as well as several low-frequency sine terms, and subsequently subtracted this baseline. Measurement runs (2) to (3) contain saturated lines, which we masked during baseline fitting and removal. For run (4), we masked all lines because a large number of lines were saturated and a significant number of weaker lines were missing in the MARVELized ExoMol/AYTY line list. We did the same for some segments of runs (1) to (3) as well, since this yielded the smallest structure in the baseline. Finally, we interleaved the baseline-corrected spectra to obtain spectra with a point spacing of 12.5 MHz. Figure 1(a) and (b) show the interleaved spectra from runs (1) and (3), where in Figure 1(b) the saturated lines are shaded in grey for clarity. The maximum signal-to-noise ratio (SNR) is ~1000 in measurement run (1) and ~600 in measurement run (3).

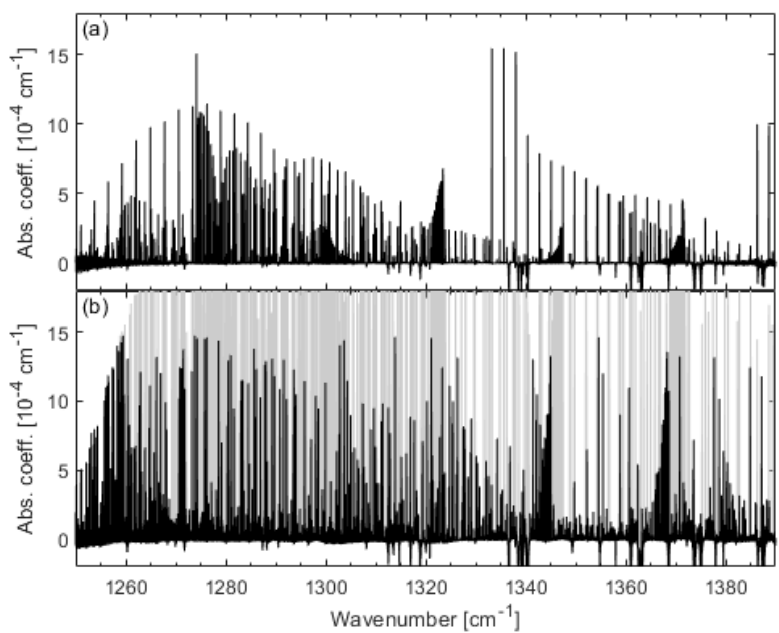

Figure 1. Interleaved absorption spectra from (a) measurement run (1) at 0.02 mbar and (b) measurement run (3) at 0.58 mbar, both averaged 400 times. In (b), the analyzed lines are shown in black, whereas the stronger lines, saturated at this pressure, are shaded in grey. Baseline distortions (negative excursions) resulting from $H_2O$ absorption in the ambient air are visible at a few positions in each panel.

## 3 Determination of experimental line parameters

### 3.1 *Line-by-line fitting and line assignment*

To retrieve line parameters from the recorded spectra, we fitted Voigt profiles to the individual observed absorption features. The positions of many lines are predicted by the MARVEL formalism[25], and hence known with uncertainties on the order of 10 MHz, while for other lines, the positions were initially only known from the semi-empirical ExoMol/AYTY line list from Al-Refaie et al.[7] and thus with much larger uncertainties (on the order of a few GHz). Therefore, we treated the two sets of lines differently, as described below.

### 3.1.1 MARVEL-predicted lines

We first selected those lines from the MARVELized ExoMol/AYTY line list, whose positions are within the range of our spectra and whose intensities correspond to an SNR of at least 20 in the analyzed spectrum. For each of these lines, we fitted a Voigt line profile to an interval of the spectrum centered at the predicted line position and spanning a range of ±400 MHz, i.e., about four times the Doppler full width at half maximum. Lines separated by less than 1.5 × 400 MHz were fitted together, while unresolved lines separated by less than 200 MHz were excluded from fitting. Spectral regions with saturated absorption features were also excluded. Lines with intensities corresponding to an SNR below 20, but above the noise floor and located within the fit window of another, stronger line that was fitted, were included into the fit model, but their line parameters were kept fixed to the values from the MARVELized ExoMol/AYTY line list[25].

We treated the line center position, the line intensity, and the Lorentzian width as free parameters, while holding the Doppler width fixed at the value calculated for the temperature of the respective measurement run (see Table 1). As initial values for the fitting procedure, we used the line positions and intensities retrieved from the MARVELized ExoMol/AYTY line list [25] together with an empirical value for the Lorentzian width of 2 MHz. The Lorentzian width contains a contribution from the pressure broadening and an instrumental broadening, the latter possibly caused by the MIR beam divergence and wavefront distortions caused by the multi-pass absorption cell (see also discussion in Ref. [55]). Figure 2 shows an example of a fitted line profile from measurement run (1) at 0.02 mbar, demonstrating structureless fit residuals.

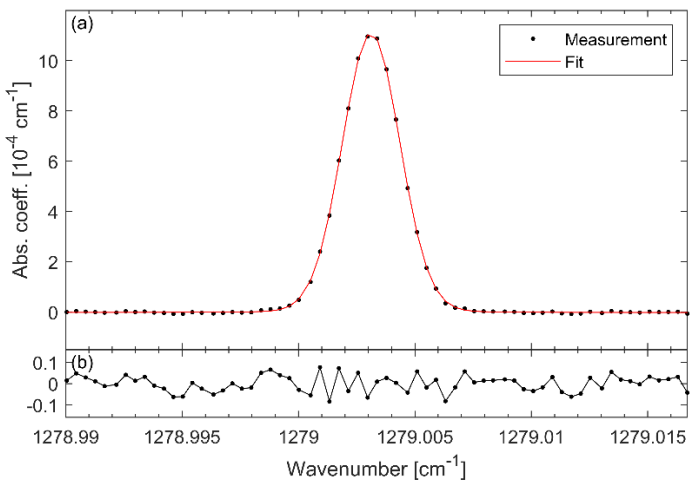

Figure 2: (a) Measured absorption due to the ($J$, $K_a$, $K_c$): (14, 0, 14) ← (13, 1, 13) line from the $\nu_6$ formaldehyde fundamental band at 0.02 mbar (black markers) together with a fitted Voigt line profile (red line). (b) Fit residuals.

### 3.1.2 Lines without MARVEL prediction

#### 3.1.2.1 Location of absorption peaks and line fitting

The recorded spectra contain several hundred absorption lines that cannot be assigned to one of the lines from the MARVELized ExoMol/AYTY line list[25]. To fit and assign these lines, we started by finding the positions and heights of absorption peaks using MATLAB's 'findpeaks' function. To convert the peak heights (in units of the absorption coefficient) to approximate line intensities, we multiplied them with a factor found from the ratio of the maximum absorption coefficient to the line intensity of a (previously assigned) line listed in the MARVELized ExoMol/AYTY line list. From the resulting list of approximate line positions and intensities, we then selected the lines that are located neither at the position of MARVEL-predicted $H_2CO$ lines, nor at those of water absorption lines (as listed in the HITRAN2020 database[29]). Using these approximate line positions and intensities as initial values, we then fitted Voigt line profiles to the observed absorption peaks as described above for the MARVEL-predicted lines.

### *3.1.2.2 Line assignment*

To predict the line positions with an accuracy sufficient for line assignment, we simulated the formaldehyde spectrum using the program PGOPHER[53]. We set up an effective rotational-vibrational Hamiltonian for the (coupled) $\nu_3$, $\nu_4$, and $\nu_6$ formaldehyde bands based on molecular $H_2CO$ constants from Refs. [52, 64]. To improve the agreement between the Hamiltonian-predicted and observed line positions, we imported into PGOPHER the line positions and assignments of all the lines from these three bands that are listed in the MARVELized ExoMol/AYTY line list with line-position uncertainties below 40 MHz and re-fitted the Hamiltonian constants. Using this refined Hamiltonian, we were able to reproduce our experimental spectra with line position deviations on the order of a few hundred MHz. For reference, we provide the PGOPHER simulation file as Electronic Supplementary Information. Combining the line positions predicted by our Hamiltonian with the *ab initio* line intensities from the semi-empirical ExoMol/AYTY line list, we compiled an augmented theoretical line list of the $\nu_3$, $\nu_4$, and $\nu_6$ fundamental bands. For the $2\nu_6 - \nu_6$ hot band, we derived the upper-state energies from the HITRAN2020[29] line list of the $2\nu_6$ overtone band (originating from the study of Perrin et al. [65]) from which we subtracted the lower-state energies predicted by our Hamiltonian to obtain the line positions of the $2\nu_6 - \nu_6$ hot band. These positions we again combined with the intensities from the semi-empirical ExoMol/AYTY line list.

To assign the fitted lines in the measured spectra, we imported the line positions predicted using the refined PGOPHER Hamiltonian into MATLAB. We compared each fitted peak with neighboring lines from our augmented theoretical line list in a two-dimensional space spanned by the line positions and line intensities. We normalized the difference of the line position and the line intensity between the peak to be assigned and each nearby line from our theoretical line list by their uncertainties. We then assigned the fitted peak to that line from the theoretical line list with the shortest Euclidian distance in this rescaled position/intensity space. If two fitted peaks were initially assigned to the same line from the theoretical line list, we reassigned that one which matched worse to the next best matching entry from our theoretical line list. Finally, we discarded lines that deviated more than 600 MHz from their predicted line position.

Using the above procedure, we assigned about a quarter of the peaks to which we could fit line profiles. The remaining ones are likely hot-band lines, whose upper states cannot be found among the corresponding overtone-band lines in HITRAN. For the $2\nu_6 - \nu_6$ band, e.g., the semi-empirical ExoMol/AYTY line list contains several dozen lines below 1300 cm$^{-1}$ with intensities on the order of those observed but whose upper states are not found in the HITRAN line list of the $2\nu_6$ overtone band. Consequently, only lines above 1300 cm$^{-1}$ could be assigned for this band. Likewise, the semi-empirical ExoMol/AYTY line list contains many lines from other hot bands (such as $\nu_4 + \nu_6 - \nu_4$, $2\nu_4 - \nu_6$, and $\nu_4 + \nu_6 - \nu_6$) with line positions and intensities in the relevant range, but for the majority of them the upper-state energies cannot be derived from HITRAN data in the way described above. Furthermore, some of the unassigned lines might belong to other formaldehyde isotopologues. The $H_2^{13}C^{16}O$ and $H_2^{12}C^{18}O$ isotopologues have relative abundances of 1.1% and 0.2%, respectively. The most intense $\nu_6$ and $\nu_3$ lines of the most abundant isotopologue ($H_2^{12}C^{16}O$) have intensities up to about $10^{-20}$ cm/molecule, which implies line intensities from $10^{-22}$ to $10^{-23}$ cm/molecule for these two rare isotopologues. As we have observed lines with intensities as low as $10^{-24}$ cm/molecule, lines from these rare isotopologues should be observable if they are located within the range of our spectra.

## 3.2 Uncertainty estimation and averaging of repeated measurements

### 3.2.1 Line positions

We identify two sources of uncertainty in the retrieved line positions. The first is the uncertainty of the line center of the fitted Voigt line profile. This uncertainty contribution is specific to each line and ranges from 0.07 to 4 MHz, depending strongly on the SNR for each line in a particular measurement run. The second contribution stems from the uncertainty in the determination of the FTS reference laser wavelength $\lambda_{\text{ref}}$. This uncertainty contribution is common for lines from each of the segments used for $\lambda_{\text{ref}}$ determination (see Sec. 2.3). We estimated this uncertainty to be between 0.05 to 0.6 MHz based on the spread of the optimum $\lambda_{\text{ref}}$ values determined for each segment, as described in Sec. 5.1 of Ref. [55]. The two uncertainty contributions were summed in quadrature for each retrieved line position from each of the measurement runs.

About half of the lines were measured in two or three of the four measurement runs. For these lines, we averaged the line positions obtained from each measurement weighted by the inverse of the squared uncertainty of each value and estimated the uncertainty of the resulting weighted mean by propagating the uncertainties of the individual measurements accordingly.

The reported line positions are not corrected for pressure shifts. Those, however, are likely insignificant (or barely significant) at the estimated line position uncertainties.

### 3.2.2 Line intensities

The uncertainties of the line intensities are dominated by the uncertainty of the formaldehyde density in the multi-pass absorption cell. The density depends on the purity of the formaldehyde sample with which the cell is initially filled. It can also vary while the spectra are recorded. As mentioned above, monomeric formaldehyde may polymerize and/or adsorb to surfaces, resulting in a decrease of the $H_2CO$ density in the gas sample. To detect such a decrease, we constantly monitored the total pressure in the gas cell while recording the spectra.

Another process affecting the total pressure is the leakage of ambient air into the cell. The manufacturer specifies a leak rate of $<10^{-5}$ Torr L s$^{-1}$ [66]. Our own measurement of the pressure increase of the sealed, previously evacuated cell yielded a value roughly one order of magnitude below that specification. Thus, the leak-rate-induced pressure increase at a volume of 0.9 L of the gas cell and the connected tubing is insignificant for measurement run (4) at 2.47 mbar (given the resolution of the pressure measurement) but relevant for runs (1) to (3) at lower pressure.

Since the leak rate was insignificant for measurement run (4), and because of the additional precautions taken to reduce contamination and dilution of $H_2CO$ during pyrolysis (see Sec. 2.2), we regard the measured line intensities from the last measurement most trustworthy. Based on the pressure measurements from that run and assuming an initial $H_2CO$ concentration of 100%, we thus calculated absolute line intensities from the fitted integrated absorption coefficients of the 151 MARVEL-predicted lines observed in run (4). We then used the MARVEL-predicted lines that have been measured both in run (3) and (4) (54 lines) to re-scale the absolute intensities from all the lines measured in run (3). To do this, we multiplied the line intensities obtained from run (3) by a correction factor such that the average ratio of the intensities from run (3) and run (4) is one. Similarly, we used the common subset of lines measured in run (2) and (3) (86 lines) to re-scale the absolute intensities of all the lines measured in run (2), and finally the common subset of run (1) and (2) (179 lines) to re-scale the intensities from run (1).

For lines measured in two or three of the four measurement runs, we report the mean of all individual measurements in the experimental line list, each of them calibrated as described.

As a crosscheck, we re-analyzed the line intensities from run (1), the measurement at the lowest pressure, with an alternative method. For that run, the two processes affecting the total pressure in the multi-pass cell can be approximately separated. At the beginning of the measurement, the total pressure in the cell decreased, which we attribute to polymerization of formaldehyde and/or adsorption at surfaces. Towards the end of the measurement, the total pressure increased because of ambient air leaking into the cell. To determine the formaldehyde partial pressure during the acquisition of the absorption spectrum, we fitted a linear model to the last hour of the recorded pressure values to determine the leak rate of the gas cell, assuming that the adsorption/polymerization processes have decreased to an insignificant level by that time. We then subtracted the increase of the total pressure due to this leak rate from the total pressure measured while recording the spectrum. The time average of this corrected pressure gives an effective formaldehyde pressure. The line intensities obtained from measurement run (1) using this effective pressure are 0.4% higher than those obtained with the rescaling based on common lines between measurement runs. The mean relative uncertainty of the line intensities from run (1) obtained from line profile fitting is 1.5%. Therefore, the line intensities from both analysis methods agree within their uncertainties.

As evident from the description above, the observed line intensities are affected by several factors with uncertainties that are difficult to quantify. Comparing the observed intensities of lines measured in two or three of the four measurement runs, we observe deviations of typically 5 to 9% (after applying the above-mentioned rescaling). Those, however, do not account for systematic deviations common to several measurement runs, such as impure formaldehyde samples. Therefore, we conservatively estimate the total line intensity uncertainties to about 15%.

## 4 Experimental line lists

The experimental line lists are available as Electronic Supplementary Information. For each line, we list the observed line center position with the corresponding uncertainty and the predicted line position retrieved from the literature (either from the MARVELized ExoMol/AYTY line list[25] or from the semi-empirical ExoMol/AYTY line list[7]). For lines assigned with our Hamiltonian model, we also give the Hamiltonian-predicted line position used for line assignment. Furthermore, we list the observed and the calculated line intensity (retrieved from the semi-empirical ExoMol/AYTY line list[7]) and the line assignment, i.e., the vibrational normal mode quantum numbers ($v_1$ to $v_6$) and the asymmetric spherical top quantum numbers ($J$, $K_a$, $K_c$) of the upper and the lower state. Table 2 summarizes key figures of the empirical line lists separated by vibrational bands.[b]

### *4.1  Line positions*

The observed positions of the MARVEL-predicted lines are shown in Figure 3 relative to the predictions from the MARVELized ExoMol/AYTY line list[25]. The rms deviations between the observed and predicted line positions are between 10 and 30 MHz, as can also be seen from Table 2. The fraction of lines with deviations below one combined standard uncertainty (1σ) of the measurement and the MARVEL prediction (summed in quadrature) are 64%, 50% and 67% for the $v_4$, $2v_6 - v_6$ and $2v_6 - v_4$ band, respectively. For the $v_6$ band, only 19% of the observed line positions agree on a 1σ level, while none of the $v_3$, $v_4+v_6 - v_4$ and $v_3+v_6 - v_3$ lines show such an agreement. As can be seen in Table 2, the observed line positions from these four bands systematically deviate from the MARVEL predictions (mean and rms deviations are of similar magnitude).

Figure 4 shows the observed positions of the lines not predicted by the MARVEL formalism relative to the values from the semi-empirical ExoMol/AYTY line list [7]. As expected, the deviations are considerably larger than those shown in Figure 3, reflecting the significantly lower accuracy of the line positions from the semi-empirical ExoMol/AYTY line list as compared to those from the MARVELized ExoMol/AYTY line list (see also Table 2). The median uncertainties of the line positions from the semi-empirical ExoMol/AYTY line list are 174 GHz (5.81 cm$^{-1}$), 157 GHz (5.23 cm$^{-1}$) and 51 GHz (1.7 cm$^{-1}$) for the lines observed here from the $v_6$, $v_4$, and $2v_6 - v_6$ band, respectively (calculated according to the expression given in Sec. 5 of Ref. [25]). Because of these large uncertainties, all observed line positions agree with the theoretical predictions on a 1σ level.

---

[b] Comparison of the MARVEL-derived line positions with those predicted by our Hamiltonian revealed that two rovibrational levels assigned to the $v_6$ vibrational state in the MARVELized ExoMol/AYTY line list [25] should better be assigned to the $v_4$ state. For another two $v_6$ levels, the $K_a$ and $K_c$ quantum numbers were reassigned. The tables and figures show the results after these reassignments. (Since the Hamiltonian has elements off-diagonal in $K_a$, $K_c$ and the vibrational quantum numbers, these are only "near" quantum numbers. Assignments of energy levels to these quantum numbers are thus only approximate and depend on the actual Hamiltonian parameters; see, e.g., [67].)

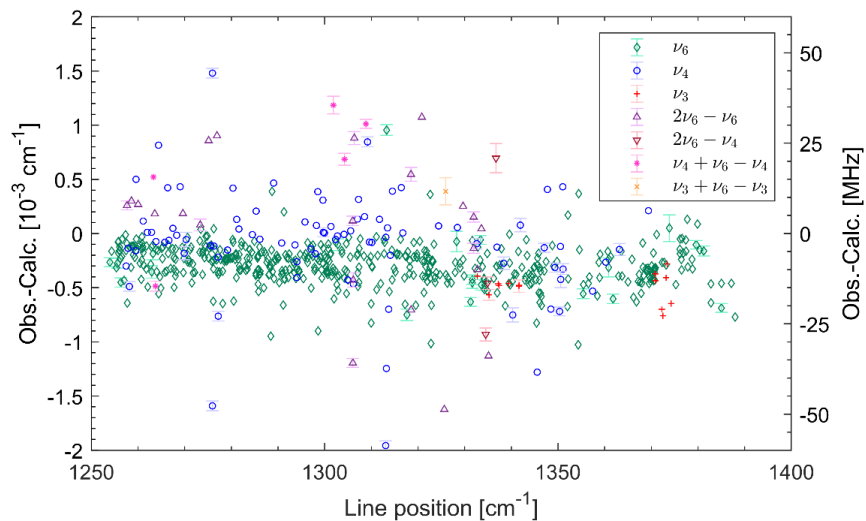

Figure 3: Observed line positions relative to MARVEL predictions for MARVEL-predicted lines. The error bars show the uncertainties of the observed positions. For clarity, error bars smaller than the symbols (i.e., < 0.04 × 10$^{-3}$ cm$^{-1}$ / 1.2 MHz) are not shown.

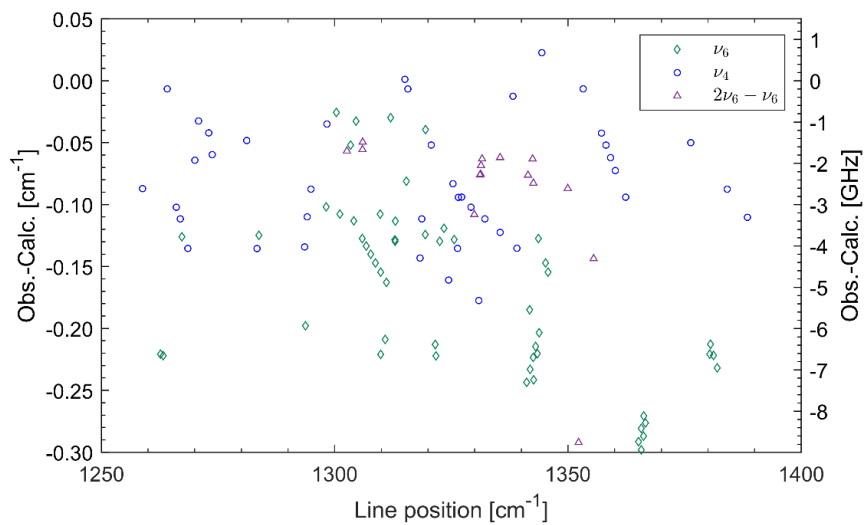

Figure 4: Observed line positions of lines not predicted by the MARVEL formalism relative to the values from the semi-empirical ExoMol/AYTY line list. The error bars of the observed positions are all smaller than the symbols shown and therefore omitted.

Table 2: Key figures of the experimental line lists. Columns: 'Band', vibrational band designation. 'Lines analyzed', number of lines analyzed from the respective band. The figures in parentheses give the number of lines with / without MARVEL-predicted line positions for the $\nu_6$, $\nu_4$ and $2\nu_6 - \nu_6$ bands. For the other bands, only MARVEL-predicted lines have been analyzed. 'Line position uncertainties', typical (i.e., median) uncertainties and uncertainty ranges of observed line positions. 'Line position deviation', mean and root-mean-square deviation of observed line positions from literature values, i.e., from those given in the MARVELized ExoMol/AYTY line list (first and single numbers) or from the semi-empirical ExoMol/AYTY line list (second numbers). 'Line intensity deviations', mean and root-mean-square relative deviations between observed and calculated (i.e., ExoMol/AYTY) line intensities, defined as: $\Delta S = (S_{obs} - S_{calc})/S_{calc}$.

| Band | Lines analyzed | Line position uncertainties [MHz] | | Line position deviation [MHz] | | Line intensity deviation [%] | |
|---|---|---|---|---|---|---|---|
| | | Median | Range | Mean | rms | Mean | rms |
| $\nu_6$ | 558 (505 / 53) | 0.37 | 0.12 – 3.7 | -8.7 / -5073 | 10.3 / 5503 | -13 | 17 |
| $\nu_4$ | 129 (88 / 41) | 0.70 | 0.30 – 2.8 | -2.5 / -2403 | 14.4 / 2796 | -6 | 26 |
| $\nu_3$ | 14 | 0.73 | 0.25 – 1.7 | -14.8 | 15.3 | 17 | 66 |
| $2\nu_6 - \nu_6$ | 37 (22 / 15) | 1.3 | 0.68 – 2.8 | 0.8 / -2717 | 20.6 / 3234 | 9 | 55 |
| $2\nu_6 - \nu_4$ | 3 | 1.8 | 1.7 – 4.0 | -6.9 | 21.6 | -11 | 16 |
| $\nu_4 + \nu_6 - \nu_4$ | 5 | 1.4 | 1.0 – 2.5 | 17.5 | 24.8 | -2 | 10 |
| $\nu_3 + \nu_6 - \nu_3$ | 1 | 3.7 | 3.7 – 3.7 | 11.7 | 11.7 | -61 | 61 |

## *4.2 Line intensities*

The measured line intensities are shown in Figure 5(a), for both MARVEL-predicted and non-MARVEL lines, and the corresponding residuals with respect to the calculated values from the ExoMol/AYTY line list are displayed in Figure 5(b). As can be seen from Figure 5(b) and Table 2, the measured and predicted intensities typically deviate by about 10 to 20%, with a few outliers deviating much more, some by >200%.

On average, the observed line intensities are lower than the predicted ones for all bands except $\nu_3$ and $2\nu_6 - \nu_6$. This deviation might be either because the observed intensities are too low – e.g., because of dilution of the formaldehyde in the sample gas by byproducts and/or contaminations (see Sec. 2.2 and 3.2.2) – or because the calculated ExoMol/AYTY intensities are too high. For comparison, we show in Figure 6 a zoom-in on our residuals together with the residuals of the experimental intensities reported by Nadler et al.[48] of a few fundamental-band lines, located outside of the spectral range investigated in this work, with respect to the ExoMol/AYTY values. These observed intensities show a similar pattern as our measurements. They are below the ExoMol/AYTY values for the $\nu_6$ and $\nu_4$ band, but above for the $\nu_3$ band. While we cannot exclude that the difficulties mentioned in Sec. 3.2.2 might lead to the measured intensities being too low, the observed discrepancy of the ExoMol/AYTY intensities with those reported by Nadler et al. could indicate that the deviations of our measurements from the ExoMol/AYTY values might, at least partially, be due to overestimated theoretical intensities. Indeed, intensity deviations on the order of 10 to 20% may arise due to the accuracy of the underlying dipole moment surface (DMS), which was computed by Al-Refaie et al.[7] at the CCSD(T)/aug-cc-pVQZ level of theory, and have been observed before for this DMS, see Table 7 in Ref. [7].

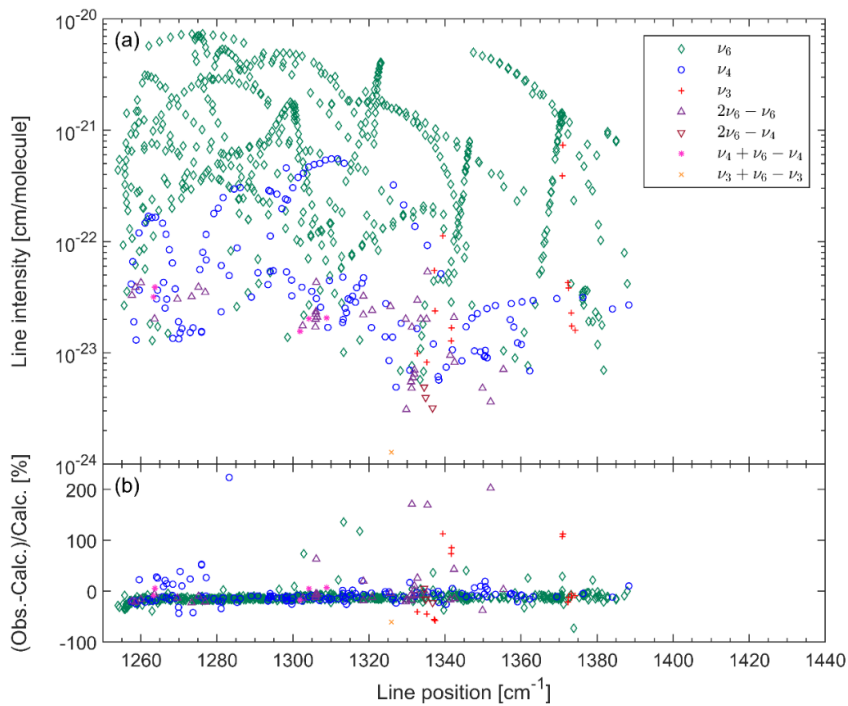

Figure 5: (a) Observed line intensities and (b) relative deviations between the observed and the calculated ExoMol/AYTY intensities.

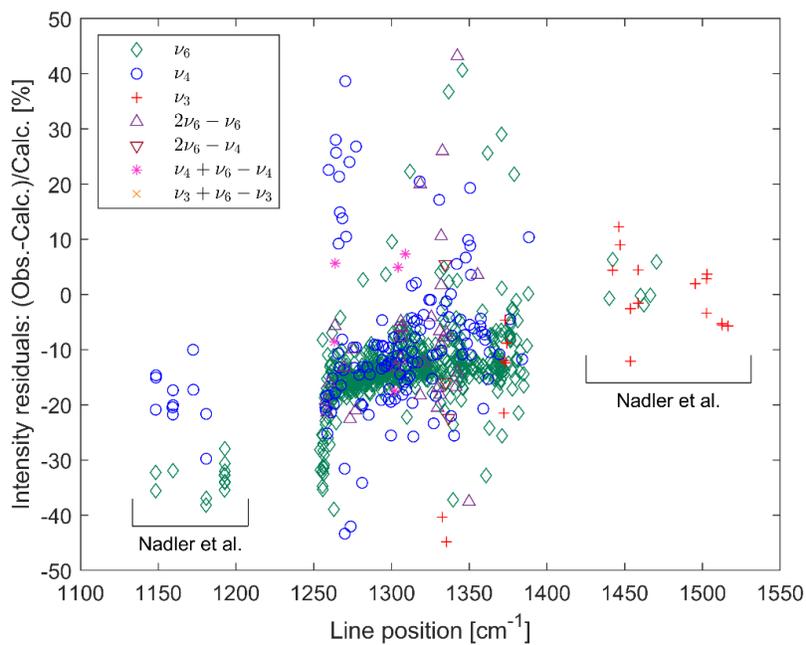

Figure 6: Zoomed view of the relative deviations between observed and calculated ExoMol/AYTY line intensities. Lines observed in this work, with relative deviations below 50%, are shown in the central portion between 1250 and 1400 cm$^{-1}$. For comparison, relative deviations between observed and calculated ExoMol/AYTY line intensities for lines reported by Nadler et al. [48] below 1200 cm$^{-1}$ and above 1400 cm$^{-1}$ are also shown.

As mentioned above, a handful of lines show considerable deviations between observed and predicted intensities, with residuals exceeding 100% or even 200%. Such deviations for only a few lines cannot be explained by an erroneous formaldehyde concentration or pressure, as those factors would affect an entire measurement run. Also, line shape distortions, e.g., due to blended lines, that could affect the fitted integrated absorption coefficient do not result in as large deviations. We visually inspected the spectra and line profile fits of all lines with residuals >100% for line blending or line-profile or baseline distortions but could not find distortions large enough to account for the observed deviations. We also looked for nearby lines predicted by our Hamiltonian that might be closely overlapping with the fitted lines or result in incorrect assignment of these lines but could not find any obvious problems. (This, however, does not entirely exclude the possibility of such occurrences because our Hamiltonian model does not predict hot-band lines.) Therefore, we attribute these large deviations between predicted and observed intensities to resonance interactions affecting the calculated intensities. It is well known that such interactions can cause large shifts in the intensities of transitions between the states involved[68, 69]; given the closeness of $\nu_6$ and $\nu_4$ vibrational bands in formaldehyde one would expect to observe resonance interactions between states from these two bands.

## 5 Revised MARVEL analysis

To fully exploit the high-accuracy line positions determined in this work and to obtain a self-consistent set of formaldehyde line positions, the positions of the 747 lines assigned in this work were added to the previous MARVEL transitions file[25] and re-run through MARVEL. This resulted in 5111 empirical energies, an increase of 82 as compared to the previously used data set. Of these 82 newly added rovibrational energies, 40 belong to the $\nu_6$ vibrational state, 28 to $\nu_4$ and 14 to $2\nu_6$. A list of all updated MARVEL-predicted energy levels is provided as Electronic Supplementary Information. The entire states file is available on the ExoMol website (www.exomol.com) as part of the recommended line list for $H_2CO$. In accordance with the current ExoMol practice[70], the updated states file also includes the MARVEL uncertainties and a tag to show which levels are computed and which come from MARVEL.

For the levels already previously included in the MARVEL analysis, adding the accurate line positions from this work resulted in revised energies with significantly lower uncertainties. In Figure 7(a), the MARVEL energies before and after adding the line positions from this work are compared. As can be seen, the new measurements propagate via spectroscopic networks to spectral regions far away from those directly probed here.

Figure 7(b) shows the reduction of the energy uncertainties. In the spectral regions probed, uncertainties decrease frequently by around one order of magnitude, some by up to two orders of magnitude. Again, the linking of empirical data through spectroscopic networks results in significantly (up to a factor of eight) reduced uncertainties in spectral regions not directly probed, i.e., up to energies of about 6000 cm$^{-1}$. Table 3 shows the median uncertainties of the rovibrational levels from the corresponding vibrational state, for vibrational states directly probed in this work, as well as those whose median uncertainty could be reduced by at least a factor of two.

The updated MARVEL energy levels were used to update the AYTY line list by replacing the calculated energy levels with MARVEL ones. Based on the newly included energy levels, the number of MARVEL-predicted line positions could be increased from 367 778 (see Ref. [25]) to 373 160. The updated MARVEL line list is provided as Electronic Supplementary Information.

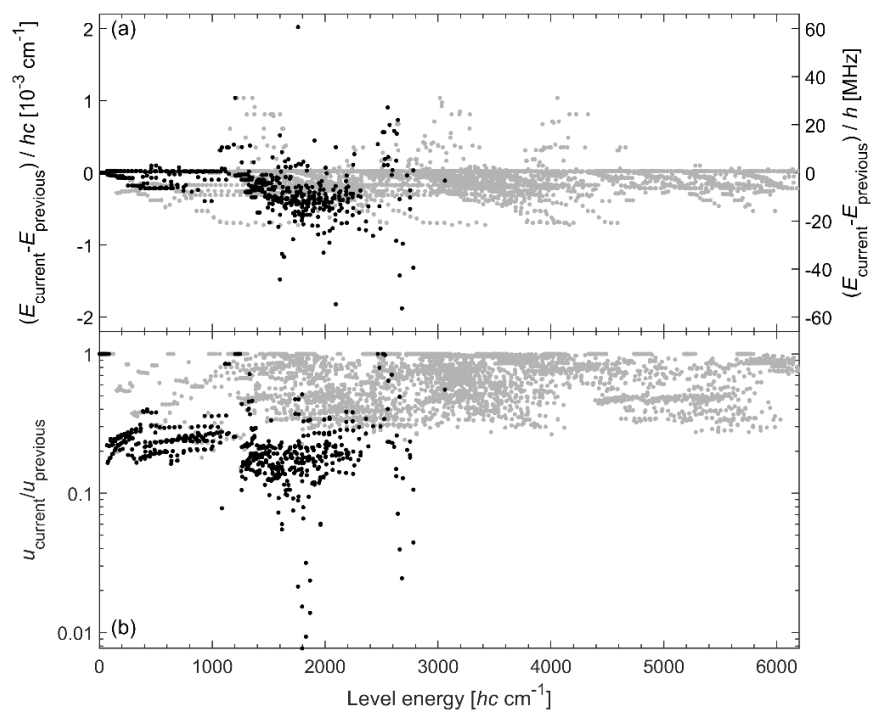

Figure 7: Impact of the high-accuracy formaldehyde line positions on the MARVEL-derived energy levels. (a) Level shifts, defined as the difference between the energy before and after revising the MARVEL analysis with the line positions from this work. The revised energies of the levels directly probed in this work (black data points) propagate via spectroscopic networks to a multitude of other levels (gray data points), some of them located far from the probed spectral range. (b) Ratio of energy uncertainties before and after revising the MARVEL analysis with the line positions from this work. Uncertainties of levels directly probed (black data points) are reduced by up to two orders of magnitude, while many other levels (gray data points) benefit from substantial uncertainty reductions because of their links to the probed ones through spectroscopic networks. (The symbols $h$ and $c$ denote the Planck constant and vacuum speed of light, respectively.)

Table 3: Uncertainty reduction of MARVEL-derived energy levels. Columns 2 and 3 show the median uncertainties of the rovibrational levels of each vibrational state listed in Column 1 before and after including the high-accuracy data from this work, and Column 4 shows their ratios (current/previous data). Bold font: vibrational states directly probed in this work. Regular font: vibrational states not probed, whose median uncertainties have been reduced by at least a factor of two through spectroscopic networks.

| Vibrational state | Median uncertainty, previous data [$10^{-3}$ cm$^{-1}$] | Median uncertainty, current data [$10^{-3}$ cm$^{-1}$] | Ratio (current/previous data) |
|---|---|---|---|
| **vib. ground state** | **0.64** | **0.19** | **0.30** |
| $\nu_2$ | 0.73 | 0.33 | 0.45 |
| **$\nu_3$** | **0.78** | **0.38** | **0.49** |
| **$\nu_4$** | **0.60** | **0.19** | **0.32** |
| **$\nu_6$** | **0.72** | **0.16** | **0.22** |
| $\nu_3+\nu_5$ | 0.45 | 0.22 | 0.48 |
| **$\nu_3+\nu_6$** | **0.58** | **0.37** | **0.64** |
| **$\nu_4+\nu_6$** | **0.55** | **0.53** | **0.96** |
| **$2\nu_6$** | **0.53** | **0.27** | **0.50** |
| $2\nu_2+\nu_6$ | 0.53 | 0.26 | 0.48 |
| $3\nu_2$ | 0.54 | 0.26 | 0.48 |

# 6 Summary and Conclusions

Using a MIR frequency comb and a Fourier transform spectrometer, we recorded spectra of low-pressure, room-temperature formaldehyde gas samples in the 8 µm spectral range. We fitted Voigt line profiles to over a thousand unique absorption peaks identified in these spectra, of which we could assign 747 to rovibrational formaldehyde transitions; 638 lines were assigned using the MARVELized ExoMol/AYTY line list from Al-Derzi et al.[25] and another 109 using an asymmetric top Hamiltonian fitted to empirical line positions from that line list. From the line profile fits, we retrieved positions and intensities of 701 transitions from the $\nu_3$, $\nu_4$ and $\nu_6$ fundamental bands as well as of 46 transitions from four hot bands with a median uncertainty of 0.4 MHz for line positions and 15% relative uncertainty for line intensities. For the line intensities, we found deviations from the calculated ExoMol/AYTY values on the order of 10%. Deviations of similar magnitude and with the same signs are found when comparing the ExoMol/AYTY values with the measured intensities reported by Nadler et al.[48].

We included the highly accurate transition frequencies obtained in this work into the MARVEL analysis of the $H_2CO$ rotational and rovibrational spectrum, which incorporates a large set of measured transition frequencies from previous studies of the formaldehyde spectrum. Adding the new line positions resulted in 82 new MARVEL-derived empirical energy levels and 5382 new MARVEL-predicted rovibrational line positions. The uncertainties of the energy levels directly probed in this work were reduced by up to two orders of magnitude. Thanks to spectroscopic networks connecting rovibrational energy levels, adding the accurate measured line positions also resulted in considerably reduced uncertainties of levels not directly probed here, ranging from vibrational ground-state rotational levels below 100 cm$^{-1}$ up to triply vibrationally excited levels at 6000 cm$^{-1}$.

Our high-resolution study also identified some deficiencies in the computed semi-empirical ExoMol/AYTY line list for hot formaldehyde[7]. Recently Mellor et al. have conducted rather similar studies for the analogous thioformaldehyde ($H_2CS$) molecule[71, 72]. A particular feature of this work was the use of a newly developed procedure based on an exact nuclear-motion kinetic energy operator, which led to the determination of a significantly more accurate semi-empirical potential energy surface and hence the calculation of a more reliable line list. Use of this improved theoretical procedure for formaldehyde should lead to an improved line list which does not suffer from the defects identified with the semi-empirical ExoMol/AYTY line list in this study.

This work lays the foundation for the addition of the bands analyzed here to the HITRAN database. While the MARVEL-derived line positions and the experimental intensities can be used straightforwardly, more work needs to be done to provide a complete (at least at terrestrial temperatures) line list. Future work will thus include a) investigating the possibility of scaling the *ab initio* intensities based on our observations, b) the use of the Hamiltonian for predicting the positions of the unobserved transitions that cannot be predicted using the MARVEL procedure, c) adding broadening parameters, and d) validating the line list against the PNNL spectra. For adding the self- and air- broadening parameters, one can use the data from Jacquemart et al.[73] when assuming negligible vibrational dependence. Similarly, the estimates of $H_2$ and He broadening reported by Tan et al.[74] can be used.

The combination of DFG-based frequency comb Fourier transform spectroscopy and MARVEL analysis used here can be applied to study mid-infrared spectra of other molecules of interest to astrophysics or atmospheric physics that absorb in the 8 µm region and are amenable to MARVEL analysis, such as the above-mentioned thioformaldehyde[71, 72], methanol ($CH_3OH$), ammonia ($NH_3$)[75] or sulfur dioxide ($SO_2$). This opens up for significant improvements of the accuracy of the line list of those molecules.

## Author Contributions
**Matthias Germann:** Conceptualization, Data curation, Formal analysis, Investigation, Methodology, Validation, Visualization, Writing – original draft; **Adrian Hjältén**: Data curation, Formal analysis, Investigation, Methodology, Visualization, Writing – original draft; **Jonathan Tennyson**: Formal analysis, Funding acquisition, Writing – original draft; **Sergey N. Yurchenko**: Formal analysis; **Iouli E. Gordon**: Conceptualization, Methodology, Writing – review & editing; **Christian Pett, Isak Silander**, **Arkadiusz Hudzikowski, Aleksander Głuszek**: Resources; **Karol Krzempek:** Resources, Writing – review & editing; **Grzegorz Soboń**: Funding acquisition, Resources, Supervision, Writing – review & editing; **Aleksandra Foltynowicz**: Conceptualization, Funding acquisition, Methodology, Project administration, Resources, Supervision, Writing – original draft.

## Conflicts of interest
There are no conflicts to declare.

## Acknowledgments
The authors would like to thank Thorlabs Sweden for the loan of the multi-pass cell.

## Supplementary information
Supplementary information for this preprint are available from the corresponding author.

**Funding**

AF acknowledges support from the Knut and Alice Wallenberg Foundation (KAW 2015.0159 and KAW 2020.0303) and the Swedish Research Council (2016-03593 and 2020-00238). GS acknowledges support from the Foundation for Polish Science (POIR.04.04.00-00-434D/17). JT and SNY acknowledge support from the European Research Council (ERC) under the European Union's Horizon 2020 research and innovation programme through Advanced Grant number 883830 and STFC Projects No. ST/M001334/1 and ST/R000476/1. IEG's contribution was supported through the NASA grant 80NSSC20K0962.